\documentclass{aa}  
\usepackage{graphicx}
\usepackage{ulem}
\begin{document}
   \title{Why did Comet 17P/Holmes burst out?}

   \subtitle{Nucleus splitting or delayed sublimation?}

   \author{ W. J. Altenhoff\inst{1}, E. Kreysa\inst{1}, K. M.
     Menten\inst{1}, A. Sievers\inst{3}, C. Thum\inst{2},
              and A. Weiss\inst{1}
          }

   \offprints{W. J. Altenhoff}

   \institute{Max-Planck-Institut f\"ur Radioastronomie, Auf dem H\"ugel 69,
             53121 Bonn, Germany\\
              \email{author@mpifr-bonn.mpg.de}
          \and
             IRAM, University campus, F 38406 St. Martin d'Her\`es, France\\
             \email{thum@iram.fr}
           \and
              IRAM, Pico Veleta, Granada, Spain\\
              \email{Sievers@iram.es}
             }

   \date{Received     ; accepted }

   \abstract{ Based on millimeter-wavelength 
continuum observations we suggest
that the recent ``spectacle'' of comet 17P/Holmes
can be explained by a thick, air-tight
dust cover and the effects of H$_2$O sublimation, which started when the comet
arrived at  the heliocentric distance 
$\le$ 2.5 AU. The porous
structure inside the nucleus provided enough surface for additional
sublimation, which eventually led to the break up of the dust cover and
to the observed outburst. The magnitude of the particle burst can be explained
by the energy provided by insolation, stored in the dust cover and the nucleus
within the months before the outburst: the subliming surface
within the  nucleus is more than one order of magnitude larger 
than the geometric surface  of the nucleus -- 
possibly an indication of the latter's porous structure.
Another surprise is  that the abundance ratios of
several molecular species with respect to  
H$_2$O are variable.
During this apparition, comet Holmes lost about 3$\%$ of its mass,
corresponding to a ``dirty ice'' layer of 20 m.

}

  \authorrunning{Altenhoff et al.}
  \titlerunning{Comet Holmes}

  \maketitle
%

\section{Introduction}
Comet 17P/Holmes was serendipitously discovered during an outbreak on
1892 November 6 by Holmes (reported by Plummer \cite{plu}) while he was
observing the nearby Andromeda galaxy (M31).
Until early 1893 January, the comet faded from magnitude 4 to 9--10, 
after which a second eruption to $\approx$ 5 mag occurred.
Obviously, this  light curve is different in time dependence and amplitude
from what was observed during the  most recent 
apparition (2007/8). After the early
observations, Comet Holmes was lost for some years, but later on
recovered as an almost 
``dead'' comet with magnitudes of $\approx$ 16--17
near every perihelion.
Whipple (\cite{whi})  analyzed the historic data again and 
explained the two outbursts by
grazing encounters of a small  hypothetical satellite with 
the nucleus: the first one
on 1892 Nov. 4.6, 1892 and the second on 1893 January 16.3.;   
even though these
encounters could not be confirmed, his review of the historic 
observations  allows this event to be 
discussed again  in connection with 
the latest outburst discussed in this paper.

Montalto et al. (\cite{mon}) report a significant disassembly
of the nucleus, not even excluding a complete disintegration.
Earlier, Sekanina  (\cite{sek1}) had classified  types of splitting comets:
(a) single comets that break up into two or more, 
(b) comets that disintegrate or suddenly disappear, 
(c) and those with a pancake-shaped companion nucleus that 
disintegrates into microscopic dust grains.
Recently, Sekanina (\cite{sek3}) summarized the optical observations 
of 17P/Holmes
and some other comets for comparison. All types of splitting comets
start in   his  hypothesis  with a major outburst. The ``megaburst'' of 
17P/Holmes is of type (c),
starting with an exothermic reaction, resulting in a rapidly expanding
cloud of microscopic dust particles.
But not all major outbursts end in splitting: e.g.,
the one of comet Halley on 1991 February 12 at a heliocentric 
distance, $r$, of  14 AU (Sekanina et al. \cite{sek4}).

\section{The nucleus}

 \subsection{Time line} 
The ``engine'' behind the cometary activity of Comet Holmes is the  
production of gaseous water
as described by Delsemme (\cite{del}). Its
production rate, $Q$(H$_2$O), is a function
of heliocentric distance.  It is $\propto$ $1/r^2$
for low values of $r$,
while for $r$ $\ge1.5$ AU the dependence becomes
highly nonlinear.  Delsemme defines the limit of sublimation $r_0$, 
the heliocentric distance beyond which
97.5\% of the energy received by insolation is re-radiated, 
and only $\le$ 2.5 \% are used for vaporization. For water ice, 
$r_0$ is about  $2.5$ AU. 

Comet Holmes is a short-period comet in the Jupiter family (a JFC). 
Its average perihelion distance, $q$, over the last
 6 apparitions was $\approx$ 2.2 AU (Marsden and Williams \cite{mar}),
close to the limit of H$_2$O sublimation. But the perihelion
distance of the most recent apparition was at a  $q$ of $2.05$ AU.
With the steepened production rate, mentioned
above, the H$_2$O production is increased by a factor of 2.   
This is possibly responsible for the outburst. \\

The outburst happened  according to Hsieh et al. (\cite{hsi}) 
on 2007 October 23.8, 
173 days after perihelion passage, or 361 days after crossing $r$ = $2.5$ AU. 
Probably -- because of the low-level cometary activity -- the nuclear surface
was free of ice and the icy nucleus was covered by  some sort
of a rubble pile (Jewitt, \cite{jew1}) or dust-particle mantle, causing
the  delay of  visible cometary sublimation by months.
During this period, the dust cover was ``air tight'', preventing the 
sublimated gas to escape. Sublimation inside the nucleus continued until 
the gas set free by this process broke up the dust mantle -- the ``outburst''.  

\subsection{Model parameter}

\textit{Dust cover.} All accurately measured cometary nuclei show a geometric albedo, $p$
between 0.02 
and 0.05 (see  e.g. Jewitt \cite{jew5}), which is indicative of a dust cover.
If closely packed, this material is thought to have 
a density, $\rho$,  $\approx$ 1 g~cm$^{-3}$,  as inferred from numerous
radar observations (Harmon \cite{har1}; Harmon et al. \cite{har}).
Information from the collision caused by the  Deep Impact mission revealed that 
the nucleus of comet Tempel 1 
had a devolatilized dust cover of about 1 m, with very little H$_2$O
inside and none on the outside (Sunshine et al. \cite{sun}) -- 
identical to what we assume for 17P/Holmes.
Below the dust cover of comet Tempel 1, 
an at least 10 m thick layer of fine grained water ice particles was found, 
which appeared to be free of
refractory impurities! This ``clean''
ice may originate from repeated sublimation and deposition
inside the enclosed nucleus approaching and leaving the solar neighborhood.
It is likely that comet Holmes has a similar layer.
But for our model we neglect this detail and assume for the mass estimates
``dirty'' amorphous H$_2$O ice throughout the nucleus.

\textit{Diameter.} Until recently, the resolution of  optical
telescopes was not high enough to directly measure the nuclear diameters
of JFCs. Instead, absolute magnitudes of the nuclei were
determined and nuclear diameters were calculated, assuming a geometric
albedo p = 0.04, because observations constrain the albedo to 0.02 $<$ p  $\le$ 0.05
(see e.g. Jewitt \cite{jew5}).
For 17P/Holmes an absolute magnitude, H$_N$, of 16.6 (Tankredi \cite{tan})
was found and a median nuclear diameter, $d_N$, of 3.2 km derived within the
limits of 4.6 and 2.9 km, corresponding to the albedo range.
Meanwhile, Lamy et al. (\cite{lam})
report a diameter $d_N$ = 3.42 km, obtained by a single
snapshot by the Hubble Space Telescope (HST). We prefer this direct measurement,
even though it might need a  correction, if the nucleus is not spherical.

 \textit{Bulk density, porosity.} The bulk density  may change from
comet to comet, depending, e.g., on the outgassing history. For our model
the value $\rho$ = 0.5 g~c$m^{-3}$ was selected, derived by  
Rickman (\cite{ric}) for the comet Halley data and 
from observations  of 29 short period comets by Rickman et al. (\cite{ric1}). 
For the porosity (fraction of void volume/bulk volume)
we assume a value 0.60. This provides ample storage for sublimated molecular
gas inside the nucleus.

\textit{Equilibrium temperature.}
One needs to know the brightness temperature, $T_b$, of the nucleus and 
the dust grains 
to calculate their emission. In the absence of new data, we assume 
that they both will be close to the equilibrium temperature, $T_{eq}$. 
For a heliocentric distance of $r$ $\approx$2.45 AU and an albedo $p$ 
of 0.04, we assume $T_b$ $\approx$175 K, i.e., identical to $T_{eq}$.

\section{Observations}
\label{s:obs}              

\begin{table}    
\caption{Flux densities $S_\nu(250)$ at 250 GHz in a 11'' beam of 17P/Holmes. 
         $\Delta$ and $r$ are the comet's distance from Earth and Sun 
         at time T after the outbreak.}
\begin{tabular}{r@{\ }l r c c r@{}c@{}l l}\hline
\multicolumn{2}{c}{date} & \makebox[2em][c]{T} & $\Delta$ & $r$ &
                           \multicolumn{3}{c}{$S_\nu(250)$}     & ref.\\
      &                  & days                &    AU    &  AU &
                           \multicolumn{3}{c}{mJy}              & \\[1ex]\hline
Oct.  &27.105 & 3.3  & 1.630 & 2.447 &64.5 &$\pm$&2.8 & (1) \\
      &28.205 & 4.4  & 1.628 & 2.451 &55.5 & &2.6 & (2) \\
Nov.  &16.950 & 24.2 & 1.634 & 2.530 &16.0 & &4.3 & (3) \\
      &18.769 & 26.0 & 1.639 & 2.538 &11.6 & &0.9 & (3) \\
      &20.846 & 28.0 & 1.645 & 2.546 & 5.1 & &1.1 & (3) \\
      &23.998 & 31.2 & 1.657 & 2.558 & 5.9 & &0.7 & (3) \\
      &25.929 & 33.1 & 1.665 & 2.566 & 7.1 & &1.3 & (3) \\
      &28.142 & 35.3 & 1.675 & 2.575 & 4.5 & &1.1 & (3) \\
Dec.  &03.838 & 41.0 & 1.710 & 2.599 & 5.5 & &1.1 & (3) \\
      &18.333 & 55.5 & 1.850 & 2.670 & 4.1 & &1.5 & (3) \\
\hline
\end{tabular}
\vspace{0ex}{\scriptsize
\begin{trivlist}
\item[(1)] extrapolated from the flux of $2.3\pm0.1$ mJy observed at 88.6 GHz by 
           \makebox[4.6em][r]{Boissier} el al. (see Sect.~\ref{s:obs})
\item[(2)] extrapolated from $2.1\pm0.1$ mJy observed at 90.6 GHz (ibid.)
\item[(3)] this paper
\end{trivlist}
}
\end{table}         

The outburst of comet 17P/Holmes came at  an unfavorable moment, 
when on Pico Veleta
the MAx-Planck Millimeter BOlometer array (MAMBO) had not been 
installed on the 30m telescope;  
in Effelsberg the 9 mm wavelength receiver
was not operational at the 100m telescope; and  for the Atacama 
Pathfinder EXperiment (APEX) 12m telescope, 
the comet was below the declination limit. About 25 days later, when 
MAMBO went back into operation, we started a series of maps and 
ON/OFF observations at its effective frequency of 250 GHz,
trying to see the aftermath of the outburst. The observing and
evaluation procedures of MAMBO observations are standard routines and have 
been frequently reported; see e.g.  Altenhoff et al. (\cite{alt2}).

The results are collected in Table 1.
Prior to our measurements, the comet had already been detected  
with the Plateau de Bure
Interferometer (PdBI)  near  90 GHz by Boissier et al. (\cite{boi}, \cite{boi1}).
Their results, generously made available to us 
prior to publication, are included in our analysis. We have scaled
the 90 GHz flux densities, obtained  with a  synthesized beam of
5.7 $\times$ 7.3 arcsec, to the angular resolution of our MAMBO data 
(11 arcsec), and we extrapolated the signal to 250 GHz with the 
canonical spectral index of comets SI = 2.7, reversing
the procedure of Jewitt and Matthews (\cite{jew0}) to derive the spectral index
of comet Hale-Bopp. This method was intensively tested by  
Altenhoff et al. (\cite{alt1}).

Each stage of the optical development has an equivalent one at
millimeter (mm) wavelengths.          
The optical observations are summarized in  Sekanina (\cite{sek3}), e.g. with  
the total magnitude $m_1$ as  a function of time,  ``the light curve'' . 

The mm data are compiled in Table 1  and Fig. 1.
The extrapolation of the PdBI observations to 250 GHz is fairly 
accurate, and the
combined errors  of extrapolation and observation are indicated by
the size of the symbols. The 
small beam broadening by the comet,  reported by 
Boissier et al. (\cite{boi}) shows that the source is optically thin.
The two  data sets are interpolated, 
suggesting a signal loss of 7 \% per day.  The series of nuclear 
magnitudes $m_2$ shows a similar slope. 

In a separate paper, Altenhoff et al. (\cite{alt1}) 
show that most cometary
mm/radio light curves can be represented by the following equation:

    $$S_{\nu} = S_{\nu,0} \Delta^{-2} \times  r^{-1.7}$$

\noindent with $\Delta$ and $r$  the geocentric and heliocentric 
distances in AU, respectively. The constant $S_{\nu,0}$= 74.5 is 
derived from the last data points.

Thus the light curve is calculated and plotted
in Fig. 1. 
It is obviously a reasonable fit for the time  after day 33, 
when  insolation and dust production (determining the intensity 
of the mm radiation) are  apparently coming to 
equilibrium. For the first 30 days, this radio light curve is the 
baseline for the burst.
As a further indicator of cometary activity, we use the nuclear 
magnitude, $m_2$,
reported  with the astrometric positions  (Marsden \cite{mar1}).
These values with limited  accuracy are averaged over three days 
(typically over 100 observations) to reduce the noise.  
These data also confirm increased nuclear activity in the first 30 days.
Red circles show the  H$_2$O  production rates measured with the 
Solar Wind ANisotropy (SWAN) experiment
on the SOlar Heliospheric Observatory (SOHO) reported by Combi (\cite{com}).
This system has a beam of about one  degree, probing the water 
production of about 4 days.
This may be a crude guess, considering that we are using observing results
obtained with very different resolutions. Even though we guess that, with 
the resulting smearing, the production rate might fit even better to our
observed extended cometary activity.

Spectroscopic observations of HCN by Biver et al. (\cite{biv2}) 
at Pico Veleta and at the Caltech Submillimeter Observatory (CSO),  
and by Drahus et al. (\cite{dra},\cite{dra1}) with the  
Arizona Radio Observatory (ARO), are shown for comparison. 
The data sets are consistent which each 
other and  show  a steeper decay 
than the cometary activity described before.

  \begin{figure}
  \resizebox{\hsize}{!}{\includegraphics[angle=-90]{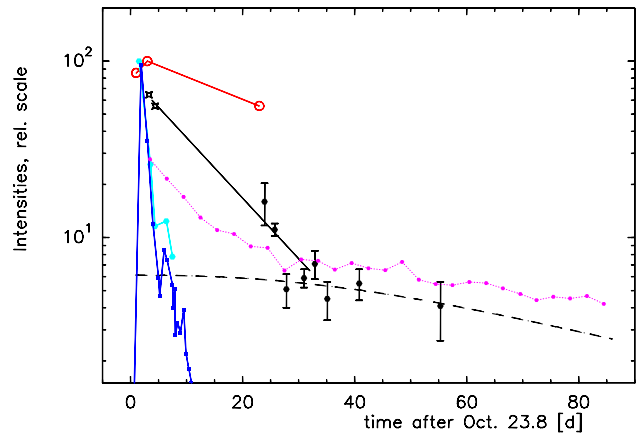}}
  \caption{Comet 17P/Holmes: Compilation of spectroscopic
      and continuum observations.       The black dots and the black 
      diamonds represent the mm continuum data at 250 GHz,
      taken with the 30m and the PdBI, respectively.
      Dashed line: model of mm halo (see text).
      Red  open circles: H$_2$O emission, observed  with SWAN,
      light blue dots: HCN
      emission, blue squares: HCN emission. See text for references.
      The dotted magenta  curve
      shows the optical nuclear magnitudes $m_2$, as an indicator of the
      nuclear activity. 
      The spectroscopic data sets are normalized to their respective maximum. }

   \end{figure}

\section{Mass determination}

\textit{Fine dust.}
Optically, the scattered light by small dust particles is dominating
the appearance of comets, even though the mass of these particles
is  low. Sekanina (\cite{sek1}) has estimated the mass of $2~\mu$m sized
fine dust in comet 17P/Holmes (see Table 2) near its outbreak. 
The size of the scattering particles 
is too small to detect with radio or mm telescopes. This dust is
responsible for the optical appearance  seen at magnitude $m_1$.
The particulate dust and the bulk of the molecular gas are almost
invisible optically.

\textit{Particulate dust.}
Radio and mm continuum observations measure the thermal emission
of dust particles of size $\ge$ 10\% of the observing wavelength, here
$\ge$ 0.2 mm. Since the observed signal is proportional to the
integrated particle cross sections, but the particle mass is
proportional to its volume, the mass of big particles
is underestimated, so 
observations at different wavelengths are needed for a more precise
mass estimate..
We estimate the dust mass with the photometric diameter to be the 
size of a  disk at the distance of the comet with its equilibrium 
temperature, radiating as black body,
yielding the same  flux density as the radio/mm halo.
For  cometary dust, we find that the black body condition 
(emissivity $\approx$1) is fulfilled
with a density of 1 g~cm$^{-3}$ and a layer depth of 3 wavelengths
(as  confirmed by the rigorous halo
evaluation for comets Hyakutake and Hale-Bopp (Altenhoff et al. \cite{alt2}).
This allows calculation of the dust mass in the halo for any observed signal.

\textit{Dust production rate.}
The average particle moves through  the telescope's diffraction beam in
about 60 hours, and the resulting dust production rate and the 
dust mass in the beam are listed in Table 2.

\textit{Hypothetical pre-burst dust.}
The radio light curve, as defined above, can be extrapolated backwards
over the whole apparition to calculate ``hypothetical'' signals 
and masses that
would have been emitted in the absence of the dust cover.
This total hypothetical mass is a factor 3 -- 4 higher 
than  the total mass released  within 33 days after  the outburst,
i.e. the burst dust mass. Thus, we can safely assume that the insolation
provided enough energy to start sublimation within the nucleus. 

\textit{Accuracy estimate.}         
Within our observing interval, the mass in the halo is approximately 
proportional to the observed flux density in the beam, thus to the 
observing accuracy. Therefore   
the  relative accuracy from day to day and of the dust production rate is 
quite good. 
The absolute accuracy depends on our knowledge of the 
absorption coefficient  $\kappa$ of cometary dust, whose uncertainty  
was estimated by Altenhoff et al. (\cite{alt2}) to be about a factor 2.  
The accuracies of the mass determinations of the small-grained dust 
(Sekanina \cite{sek}) and of water (Combi \cite{com}) have unfortunately not 
been reported.

\section{Interpretation}
\textit{Start of the outburst.}
The nuclear structure of comets 9P/Tempel 1 and 17P/Holmes before the outbursts
are probably alike, a densely packed dust cover ($\approx$1 m) below a layer
of pure water ice ($\approx$10 m), below amorphous dirty H$_2$O ice, whose upper
part is possibly free of highly volatile molecules. At 9P/Tempel 1 the impactor
acted as the exothermic energy source to blow off the pancake-shaped dust cover,
as the scheme of Sekanina (\cite{sek}) suggests, making it a type (c) 
split nucleus.  The development for 17P/Holmes is different.
When H$_2$O sublimation started inside the porous nucleus, water vapor spread
all over the nucleus, initiating more sublimation; deeper inside, and
even  other molecular ices with lower sublimation points were heated, 
stored there at lower temperatures. The effective sublimating surface 
inside the nucleus,  estimated as excess over the emission after the 
burst on day 35 when it was near equilibrium with insolation,
was more than 14 times  the nuclear surface, corresponding 
roughly to the nuclear size
of comet Hale-Bopp. The sum of the saturated partial pressures of
all ice species was obviously breaking up  the air tight dust mantle,
allowing the cometary wind to start through the dust mantle and 
lifting dust particles,
piece by piece, into the halo. The break up of the dust mantle is
hardly spectacular, compared with the full start of cometary wind.

\textit{Time scale of the outburst.}
Different versions exist of the development of the outburst.
Sekanina (\cite{sek3}, \cite{sek}) refers to an explosion and a 
single exothermic event, and
Biver et al. (\cite{biv2}) and others report a water production rate, 
which almost ends
after 3  days. In Fig. 1 the continuum observations are plotted, showing that
the outburst-related increased continuum emission lasted for about 30 days, 
as did the increased nuclear magnitude $m_2$. Additional proof 
are the numerous photographs taken within the first month of the outburst; 
see e.g. Sekanina (\cite{sek3}),
in which the comet appears as a filled Plerion rather than a shell, implying
that the dust injection into the coma continued after the ``explosion'' 
for quite some time.

\textit{Molecular production rates.}
Production of gas-phase molecules is responsible for all the cometary activity.
It is predominantly the cometary wind of the H$_2$O molecules, which
lifts the dust particles from the nucleus, so a correlation
between H$_2$O and dust production is expected. 
Usually the production of different molecules
shows a fixed ratio, so that one can, e.g., predict the H$_2$O
production rate from HCN observations. Not so for comet Holmes!
Figre 1 shows that H$_2$O production, observed with the SWAN satellite,
lasts at least for a month as does the enhanced mm continuum
emission, while e.g. the spectral lines of HCN, CO, NH$_3$
(Drahus et al. \cite{dra}; Biver et al. \cite{biv2};  Menten, \cite{men})
had a big signal at the start, which  apparently  petered out 
dramatically after 3 days,
as shown in Fig. 1. The reason may be the temperature/depth structure of
the nucleus, because the near surface ice might be free of volatile molecules.

 \textit{Mass comparison.}
All derived masses are collected in Table 2, where the total mass and
the mass of the dust cover have been calculated with the model values. 
The mass of the 2 $\mu$m sized
dust, determined immediately after the outburst by Sekanina (\cite{sek3}),
is surprisingly high, compared to the total mass of the dust layer! Even
the total particulate dust mass (grains of size 0.2 to 7 mm) is smaller. 
Integrated over the
33 days of increased cometary activity, the dust mass produced by the 
outburst is 
$\approx$ 2\% of the comet's total mass. Surprisingly low is also 
the H$_2$O mass released in the
first part of  the outburst, when we would have expected a mass 
comparable to the particulate dust
mass. The hypothetical  dust mass, calculated from the radio light 
curve backwards,
is about a factor of 3 -- 4 higher than the total mass in the burst. The energy
released in the burst can be provided by the insolation before the burst,
even allowing energy losses through re-radiation by the dust cover.

The total accounted mass loss during this apparition (mass of $2~\mu$m sized 
dust, burst dust mass, burst H$_2$O mass) is in total $\le$ 3.5\% of the 
total nuclear mass.
If a bulk density of $\rho$ = 0.5 is assumed for the outer nucleus, this
loss corresponds to a layer of 20 m thickness which is the same order
 of magnitude as found from observations of other comets.

\begin{table}
\caption{Mass budget}
\begin{tabular}{l r@{\,}l l}
\multicolumn{1}{l}{Contribution} &\multicolumn{2}{c}{Mass} & Comment     \\
\hline
nuclear mass          & $1.1\,10^{16}$ & g   & Model, $\rho$ = 0.5
                                             \rule{0em}{3ex}  \\
initial dust cover    & $3.7\,10^{13}$ & g   & Model, $\rho$ = 1.0  \\
dust ($\sim2 \mu$m)   &\ \ $1\,10^{14}$& g   & (1) \\
dust mass in halo     & $4.3\,10^{13}$ & g   & on day 3        \\
dust production rate  & $2.0\,10^{8}$  & g/s & for day 3         \\
burst dust mass       & $2.1\,10^{14}$ & g   & day 3 -- 33  \\
dust mass in halo     & $3.0\,10^{12}$ & g   & day 35 (equil.) \\
H$_2$O production rate& $3.6\,10^{7}$  & g/s & (2)   \\
burst H$_2$O mass     & $6.5\,10^{13}$ & g   & day 1 -- 23  \\
\hline
\end{tabular}
{\scriptsize \begin{list}{}{}
\item[(1)] Sekanina (\cite{sek}) 
\item[(2)] converted from $1.2\,10^{30}$ mols/s (Combi et al. \cite{com})
\end{list}}
\end{table}

\subsection{Alternative models}
Sekanina (\cite{sek},\cite{sek1}) explained the outburst of comet 17P/Holmes
as a splitting nucleus, whereby the secondary nucleus is a fragment
of a jettisoned insulation mantle of debris. The splitting starts
with an exothermic event. His model  considers neither particulate dust
with particles $\ge$ 0.2 mm and nor the H$_2$O production, both of which
contribute at least as much mass, each seperately as doeshis 
``secondary nucleus''.

\section{Conclusion}
The historic outbursts, as discussed by Whipple (\cite{whi}), show several
similarities to the present one, suggesting that they happened the 
same way, but in 2 steps.
After all, comet 17 P/Holmes is a comet like many others whose appearance is
determined by sublimation of cometary ices. What makes it peculiar is
that it had a big dust cover and that it  seldom comes close enough to the Sun
to afford a great display of activity. Dust covers of cometary 
nuclei are standard (see model of Horanyi et al. \cite{hor}) and do not
indicate a
splitting comet. We think that the delayed sublimation, as explained 
above, is a
viable alternative to the theory of splitting or sudden fragmentation of the
cometary nucleus.  
section{Acknowledgement}
We are grateful to Dr. J. Boissier (IRAM)  for communicating  the 90 GHz
results to us prior to publication.
We thank the director of IRAM, Dr. P. Cox, for granting special
observing time and the staff on Pico Veleta, Spain, for their
support of the observing program.



\begin{thebibliography}{}



  \bibitem[2008]{alt1} Altenhoff. W. J., Bertoldi, F., Thum, C.,
      et al., 2008, A\&A, in preparation

   \bibitem[2000]{alt2} Altenhoff, W. J., Bieging, J. H., Butler, B.,
        et al., 1999, A \&A, 348, 1020 - 1034

  \bibitem[2008]{biv2} Biver, N., Bockel\'ee-Morvan, D., Wiesemeyer, H., 
  et al., Asteroids, Comets, Meteors 2008 held July 14-18, 2008 in 
   Baltimore, Md.  LPI Contribution No. 1405, paper 8146 

  \bibitem[2008]{boi} Boissier, J., Bockel\'ee-Morvan, D., Biver, N., et al., 
      Asteroids, Comets, Meteors 2008 held July 14-18, 2008 in Balimore, Md.
      LPI Contribution No. 1405. paper  8081

  \bibitem[2009]{boi1} Boissier, J., Bockel\'ee-Morvan, D., Biver, N., et al.,
        2009, in preparation

  \bibitem[2007]{com} Combi. M. R., Maekinen, J. T. T., Bertaux, J.-L.
   et al., 2007, IAU Circular 8905, Central Bureau for Astronomical Telegrams,
     Cambridge, USA


  \bibitem[1982]{del} Delsemme, A. H., 1982. In: Comets, edited by 
       Wilkening, L. L., Univ. Arizona Press, Tucson, p. 85 - 130

  \bibitem[2007]{dra} Drahus, M., Paganini, L., et al., 2007, IAU Circular
     8891, Central Bureau for Astronomical Telegrams, Cambridge, USA

  \bibitem[2008]{dra1} Drahaus, M., Paganini, L., et al.,  Asteroids, 
          Comets, Meteors 2008 held July 14-18, 2008 in Baltimore, Md.
          LPI Contribution No. 1405, paper 8340 

 \bibitem[2001]{fer1} Fern\'andez, Y. R., Jewitt, D. C., \& Sheppard, S. S.,
         2001, ApJ, 553, L197 - L200

 \bibitem[2007]{gai} Gaillard, B., Lecacheux, J., and Colas, F., 2007,
     CBET 1123, Central Bureau for Astronomical Telegrams, Cambridge, USA

  \bibitem[1999]{har1} Harmon, J. K., Campbell, D. B., Ostro, S. J., \&
      Nolan, M. C., 1999, Planet. Space Sci., 47, 1409 - 1422

  \bibitem[2005]{har} Harmon, J. K., Nolan, M. C., 2005, Icarus, 176, 175 - 183

 \bibitem[1984]{hor} Horanyi, M., Gombosi, T. I., Cravens, T. E., et al., 1984,
          ApJ 278, 449 - 455

  \bibitem[2007]{hsi} Hsieh, H. H., Fitzsimmons, A., and Pollacco, D. L., 2007,
         IAU Circular 8897, Central Bureau for Astronomical Telegrams,
         Cambridge, USA

 \bibitem[1992]{jew1} Jewitt, D. C.. In: Observations and Physical Properties
     of Small Solar System Bodies, 30th Liege International Symposium,
     Liege, 1992. p. 85 - 112

 \bibitem[1999]{jew0} Jewitt, D. C., and Matthews, H., 1999, AJ 117, 1056

  \bibitem[2005]{jew5} Jewitt, D., 2005. In:  Trans-Neptunian Objects and 
   Comets, Saas-Fee Advanced Course 35. Ed.: K. Altwegg, W. Benz and N. Thomas,
       Springer-Verlag, Berlin p. 1 - 78


 \bibitem[2005]{lam} Lamy, P. L., Toth, I., Fern\'andez, Y. R., Weaver, H. A.,
      2005. In: Comets II, Ed.: Festou, H. C., Keller, H.U., and 
      Weaver, H. A., Univ.  Arizona Press, Tucson, p. 223 - 264


  \bibitem[1999]{mar} Marsden, Brian G., Williams, Gareth V.. Catalogue of
    cometary orbits 1999, 13th Edition.  Minor Planet Center, Cambridge, USA

 \bibitem[2007]{mar1} Marsden, Brian G., 2007, Observations of Comets. Bi-weekly
    listing in: Minor Planet Electronic Circulars, Minor Planet Center,
   Cambridge

  \bibitem[2007]{men} Menten, K. M., 2007, private communication

  \bibitem[2008]{mon} Montalto, M., Riffeser, A., Hopp, U., et al., 2008, 
         A\&A 479, L45 - L49

  \bibitem[1893]{plu} Plummer, W. E., Report of the Council, 1893, MN 53, 266

   \bibitem[1987]{ric1} Rickman, H., Kamel, L., Festou, M. C., and
       Froeschl\'e, C., 1987. In: Symposium on the Diversity and Similarity
       of Comets. Ed.: Rolfe, E. J., and Battrick, B.. ESA SP-278, 
       Nordwijk, p. 471 - 481

   \bibitem[1989]{ric} Rickman, H., 1989, Adv. Space Res., 9, 59 - 71

   \bibitem[2007]{sek} Sekanina, Z., 2007, Electronic Telegram 1118,
          Central Bureau for Astronomical Telegrams, Cambridge, USA

   \bibitem[1982]{sek1} Sekanina, Z., 1982. In: Comets, edited by Wilkening,
      L. L., Univ. Arizona Press, Tucson, p. 251 - 287

   \bibitem[1991]{sek2} Sekanina, Z., 1991. In: Comets in the Post-Halley
          Era, 2. Edited by Newburn, R. L., Neugebauer, M. Rahe, J., Kluwer
          Academic Publisher, Dordrecht, p. 769 - 823


   \bibitem[1992]{sek4} Sekanina, Z., Larson, S. M., Hainaut, O., at al., 1992,
           A\&A 263, 367 - 386

    \bibitem[2008]{sek3} Sekanina, Z., 2008, ICQ 30, 3 - 28


   \bibitem[2007]{sun} Sunshine, J. M., Groussin, O., Schultz, P. H., et al.,
         2007, Icarus 191, 73 - 83


   \bibitem[2006]{tan} Tancredi, G., Fernandes, J. A., Rickman, H.,
       Licandro, J., 2006, Icarus, 182, 527 - 549


  \bibitem[1986]{whi} Whipple, F., 1984, Icarus 60, 522 - 531


\end{thebibliography}
\end{document}